\documentclass[twocolumn,prl,superscriptaddress,aps]{revtex4}

\usepackage{graphicx}				% Grafiken in Abbildungen einbinden
\usepackage{amsmath}				% American Mathematical Society math function
\usepackage{amssymb}				% American Mathematical Society math symbols
\usepackage{color}
\usepackage{pdfcolmk}

\newcommand\Ef{E$_F$}
\newcommand\idx[1]{_{\textrm{\scriptsize #1}}}

\newcommand\degree{$^\circ$}
\newcommand{\comment}[1]{}

\newcommand\beq{\begin{displaymath}}
\newcommand\eeq{\end{displaymath}}
\newcommand\beqn{\begin{equation}}
\newcommand\beqnl[1]{\begin{equation}\label{#1}}
\newcommand\eeqn{\end{equation}}
\newcommand\eeqnl{\end{equation}}

\begin{document}
%\title{Generation of photostationary states of mixed isomere phases on a semimetal surface}

\title{Induction of a photostationary ring-opening/closing state of spiropyran monolayers on the semi-metallic Bi(110) surface}

\author{Gunnar Schulze, Katharina J. Franke and Jose Ignacio Pascual}

\affiliation{Institut f\"ur Experimentalphysik, Freie Universit\"at Berlin,
Arnimallee 14, 14195 Berlin, Germany}

%\date{\today}

%\begin{document}
\date{\today}
\begin{abstract}

Molecular switches on metal surfaces typically show very little photoreactivity. Using scanning tunneling microscopy we show that the ring-opening/-closing switch
nitrospiropyran thermally and optically isomerizes to the open merocyanine form on a Bi(110) surface. Irradiation by blue light of a monolayer spiropyran molecules leads to mixed domains of the two isomers. At large illumination intensities a photostationary state is established, indicating the
bidirectional ring-opening and closing reaction of these molecules on the
bismuth surface. The enhanced photoactivity contrasts with the case of
adsorption on other metal surfaces, probably due to the low density of states at the Fermi level of the semi-metallic Bi(110)
surface.
\end{abstract}

\maketitle

The combination of organic molecular switches with inorganic materials foresees
extending their switchable functionality to a large variety of physical and
chemical processes. Several interesting perspectives arise from such hybrid
systems like, for example, using the different conductance of two isomers  to
act as an electronic switch in molecular based devices \cite{Kudernac09,
vanderMolen10}, or the different optical adsorption properties to produce
tunable coatings \cite{FeringaBook, Rosario04}. The application of molecular
switches in devices, however, relies on the persistence of their switching
ability by external stimuli when interacting (mechanically and electronically)
with a metal electrode or a surface. The presence of a metal surface may
introduce alternative excitation routes in the switching mechanism:
photoexcited electrons/holes from the substrate may be transferred to the
molecule enabling an isomerization in the anionic/cationic states, respectively
\cite{Hagen08a}. In spite of that, it is found that the switching processes can
be irreversible or even fully suppressed for   molecules  in direct contact
with a metallic substrate. The origin is attributed to the fast quenching of
excited states due to the presence of the metal; the coupling of molecular
states with metallic electronic bands  allows dumping the excitation energy
into a continuum of substrate excitations. The strong reduction of excite
states' lifetime down to time scales much shorter than the isomerization
process \cite{Dulic03, Henningsen08, Tegeder11} drastically decreases the
quantum yield many orders of magnitude with respect to electronically isolated
molecules \cite{Crommie07}.

Spiropyran based molecules are prototype molecular switches. The
1,3,3-Trimethylindolino-6-nitrobenzopyrylospiran ({\em spiropyran}, SP) isomer
is a three-dimensional, inert, and colorless molecule. Cleavage of the central
C-O bond leads to the planar, chemically active, and colored {\em merocyanine}
isomer (MC) (Fig. 1). In solution, the ring-opening reaction is induced by ultra-violet
light, while the back-reaction is triggered by visible light or temperature
\cite{Rumpf53, Hirshberg52}. Multilayers of these molecules are also reversibly
photo-isomerizable \cite{Karcher07}. In contrast, monolayers of SP on a Au(111)
surface behave completely different; the only switching event observed was the
thermally activated ring-opening reaction \cite{Piantek09}. The inversion of
the thermodynamic stability and inhibition of photoreactive switching
emphasized the influence of the substrate \cite{Tegeder11}. One strategy to
reduce the role of the surface could be to employ materials which weakly couple
electronically with adsorbates.

Here, we investigate the response to illumination of spiropyran molecules
adsorbed on Bi(110), a semi-metallic surface with low density of states at the
Fermi level \cite{Hofmann01, Hofmann06}. We find that, contrary to other metal
surfaces, spiropyran exhibits here photo-reactive activity; illumination with
monochromatic (blue) light produces increasing amounts of the open isomer with
the applied power, until reaching a
 photostationary state consisting of both closed (SP) and
open (MC) isomers with a ratio reflecting their difference in
photoisomerization cross-section. The existence of this state implies that
photoisomerization occurs in both directions, i.e. the C-O bond is cleaved and
formed by the same external stimuli.

The bismuth surface was prepared by successive ion sputtering and annealing
(420 K) cycles in ultra-high vacuum. The spiropyran molecules were evaporated
using a Knudsen cell onto the bismuth sample held at room temperature. The
sample was then cooled down and transferred into a custom-made low temperature
scanning tunneling microscope (STM), at $5$~K, for inspection. For the
illumination experiments, the sample was removed from the STM  and irradiated
with a blue laser diode (wavelength $\lambda$= 445 nm, photon energy E$_{ph}$=
2.8 eV, total power P= 45 mW) in the preparation chamber with an angle of
45\degree\ and at 300 K. The samples were afterwards cooled and inspected again
at $5$~K.

\begin{figure}[tb]
  \begin{center}
    \includegraphics[width=0.95\columnwidth]{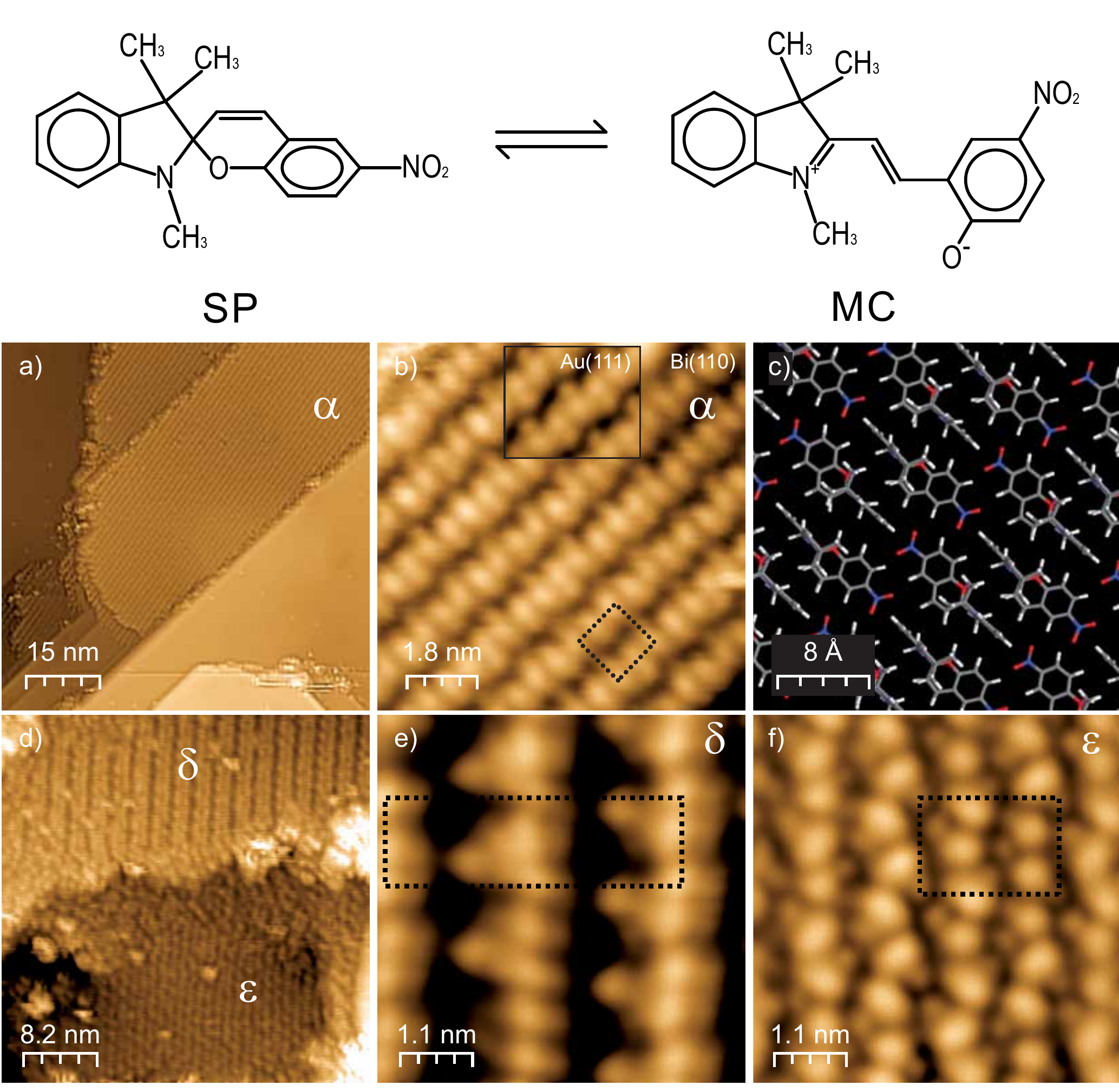}
  \end{center}
  \caption{(Color online) Chemical structure of the spiropyran (SP) and merocyanine (MC) isomer.
  (a) STM image of spiropyran islands on Bi(110) grown at room temperature and (b) its close-up view \cite{wsxm}. The inset shows the STM image of a SP island obtained on Au(111) \cite{Piantek09}, for comparison. The unit cell is
  marked by a dotted rectangle. (c) Ball and stick model of the SP island following force-field model simulations \cite{Piantek09}. (d) Overview image showing the coexistence of two molecular phases $\delta$ and $\varepsilon$ after annealing the bismuth sample to $330$~K.
  (e, f) Zoomed image of the two different phases outlining their different
  molecular structure and their unit cell.
  (a: $I\idx{t} = 50$~pA, $V\idx{bias} = 1.2$~V, b: $I\idx{t} = 50$~pA, $V\idx{bias} = 700$~mV, d: $I\idx{t} = 10$~pA, $V\idx{bias} = 1.0$~V, e: $I\idx{t} = 30$~pA, $V\idx{bias} = 1.0$~V, f: $I\idx{t} = 60$~pA, $V\idx{bias} = 1.0$~V)}
\end{figure}

Deposition of spiropyran molecules onto the Bi surface at RT leads to
self-assembled extended islands (Fig. 1 a) composed of molecular features
arranged in densely packed rows with a rectangular unit cell ($1.3 \times
1.2$~nm$^2$). The structure of the islands resembles in shape and size
spiropyran islands grown on a  Au(111) surface at temperatures below $270$~K
(see inset of Fig. 1  b) \cite{Piantek09}. From the striking resemblance of STM
images on the two surfaces we conclude that also on bismuth the ring-closed
spiropyran conformation (SP) can be stabilized on the surface.  The
corresponding structural model of the molecular arrangement is shown in Fig. 1
c). The structure is stabilized by $\pi$-H bonds and H-bonds between the
molecules. This structure is labeled as ``$\alpha$'' in the following.

The structure of the molecular layer changes drastically when the bismuth
sample is annealed to $330$~K. While the initial pattern vanishes, two new
ordered phases appear on the surface, plus some disordered regions (Fig. 1  d).
The first of the two novel phases, labeled $\delta$ (Fig. 1 e), is
characterized by broad stripes  (unit a cell $4.5 \times 1.4$~nm$^2$). Here,
the STM images do not provide any concluding fingerprint about the isomeric
composition of the structure. The second phase ($\varepsilon$) has a smaller
unit cell ($2.2 \times 1.4$~nm$^2$) and clearly shows inside a double-lobe
molecular feature repeated 4 times with different orientations (Fig. 1 f). This
homogeneous phase $\varepsilon$ is the only one observed on the sample after
annealing to higher temperatures ($350$~K). As for the case on Au(111)
\cite{Piantek09}, we expect that at sufficiently high temperatures the complete
molecular layer undergoes a thermally-activated ring-opening reaction. Hence,
we identify the high-temperature phase $\varepsilon$ as being composed
exclusively of the ring-opened isomer merocyanine (MC). Correspondingly, the
intermediate phase $\delta$ is formed by a mixture of both isomers within its
unit cell.

To investigate the photoisomerization ability of the molecules on the Bi(110)
surface, a pristine SP layer was exposed to blue laser light (E$_{ph}$= 2.8 eV)
for $240$~minutes, while keeping the sample temperature at 300 K, below the
temperature for thermal isomerization \cite{note1}. The laser spot was not
focused on the surface; instead, it had an oval shape of about
$7$~mm~$\times$~$3$~mm size. Since the photon intensity is not homogeneous
across the laser beam, we expect that different areas of the sample were
exposed to different light intensity. This allows us to investigate in a
single experiment the effect of different photon fluences on the isomerization
simply by exploring with the STM different areas within the laser spot. To
obtain a function describing the photon fluence across the surface, we
approximate the total power  distribution across the laser spot with a Gaussian
function and consider that it integrates to the calibrated value of 45 mW. In
this way, we obtain that the laser fluence during the experiment gradually
decreased from  $3.7$~kJ/cm$^2$ at the center of the spot to zero, at the
boundaries.

\begin{figure}[tb]
  \begin{center}
    \includegraphics[width=\columnwidth]{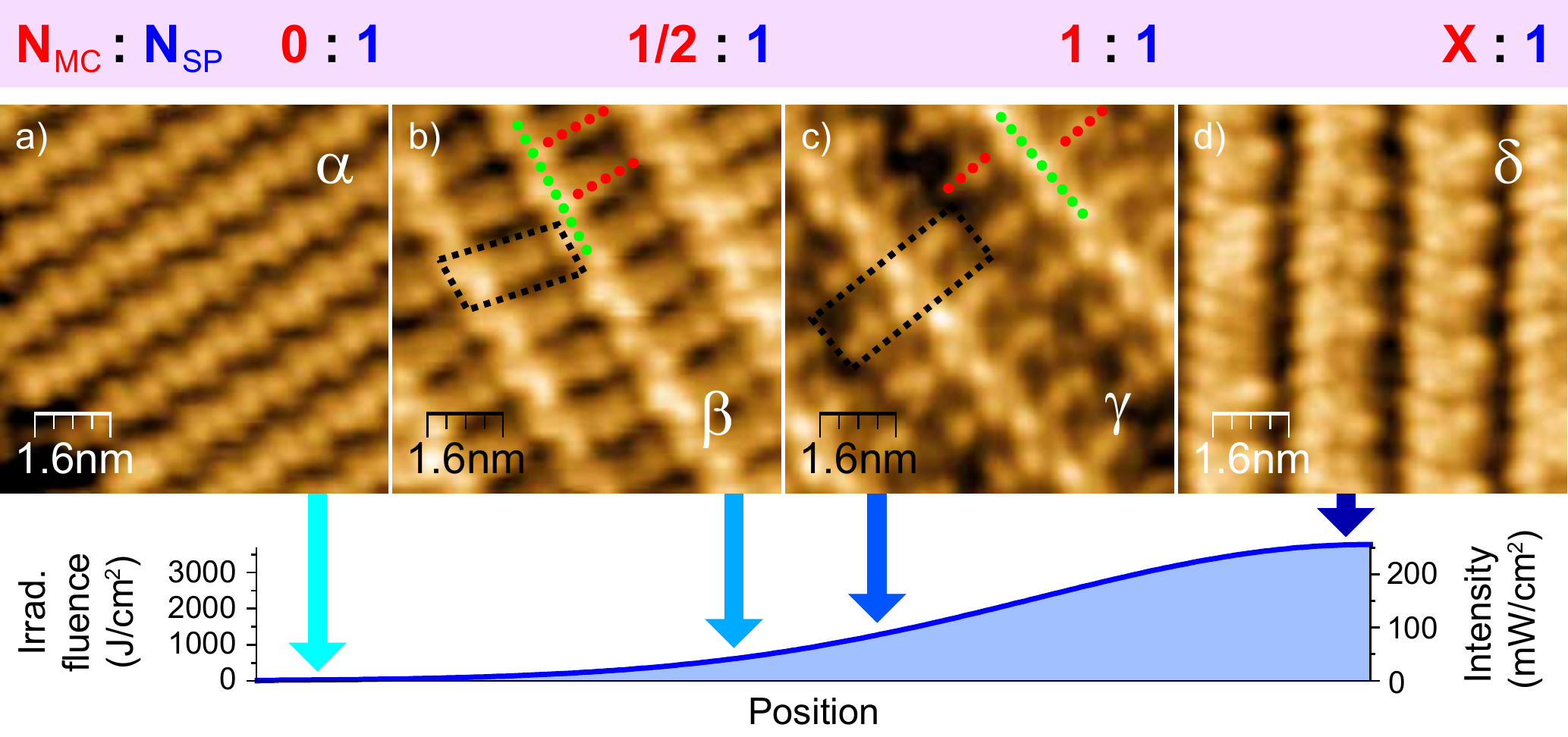}
  \end{center}
  \caption{(Color online) Photo switching activity of spiropyran on Bi(110) after 240 min
  of illumination. (a-d) show STM images of the four dominating molecular
  patterns $\alpha$, $\beta$, $\gamma$ and $\delta$ at positions with increasing radiation intensities.
  The Gaussian curve at the bottom illustrates the intensity profile. A unit cell is marked for the
  light-induced phases and the assumed MC:SP ratio of the patterns is
  noted above the images. The green dotted lines mark the structures labeled
  as ``spinal rows'' in the text while the red dotted lines highlight
  the ``ladder step'' features. After irradiating with maximal intensity the
  system saturates at structure $\delta$ (d).
  }
\end{figure}

After irradiation of the SP molecular layers, a set of new structurally
different phases appear on the surface, generally coexisting, and with
some disordered areas of molecules in between. Fig. 2 summarizes the different phases and the position where they are most frequently observed with respect to the center of the laser spot. The drastic change in molecular structure with respect to the initial $\alpha$ phase is an indication that they do not consist of pure SP molecules, but of a mixture with the other isomer (MC). The type of phases most frequently observed in a region (i.e. their relative area) depends on the position within the broad laser spot. This suggests that there is a correlation with the light distribution across the laser spot.

On positions with low irradiation power, the initial pure SP phase can still be observed (Fig. 2 a). However, in regions with moderate irradiation two novel grid phases, labelled $\beta$ and $\gamma$, appear (Fig. 2 b and 2 c). These phases combine structures with two different shapes in the STM images, forming ladder-like arrangements. Scanning tunneling spectroscopy (STS) measurements allows us to identify the corresponding isomers (Fig. 3). The spectra on the higher species show an unoccupied molecular resonance at $1.1$~V, characteristic of molecules in the $\alpha$ phase. We thus identify them as spiropyran isomers. Correspondingly, the lower features are identified as the open isomer merocyanine, with spectral features consisting of an empty state shifted to higher values and the onset of an occupied resonance below $-2.0$~V. Careful analysis of the shape and size
of the patterns' unit cell allows us to conclude molecular ratios MC:SP  of 1:2 and 1:1 for the $\beta$ and $\gamma$ phases, respectively \cite{note2}. The different composition of the two phases correlates with  different photon dosage ($0.7$~kJ/cm$^2$ for  $\beta$  and $1.3$~kJ/cm$^2$ for  $\gamma$). This is in agreement with an increase of the MC fraction for larger radiation doses.

\begin{figure}[tb]
  \begin{center}
    \includegraphics[width=0.9\columnwidth]{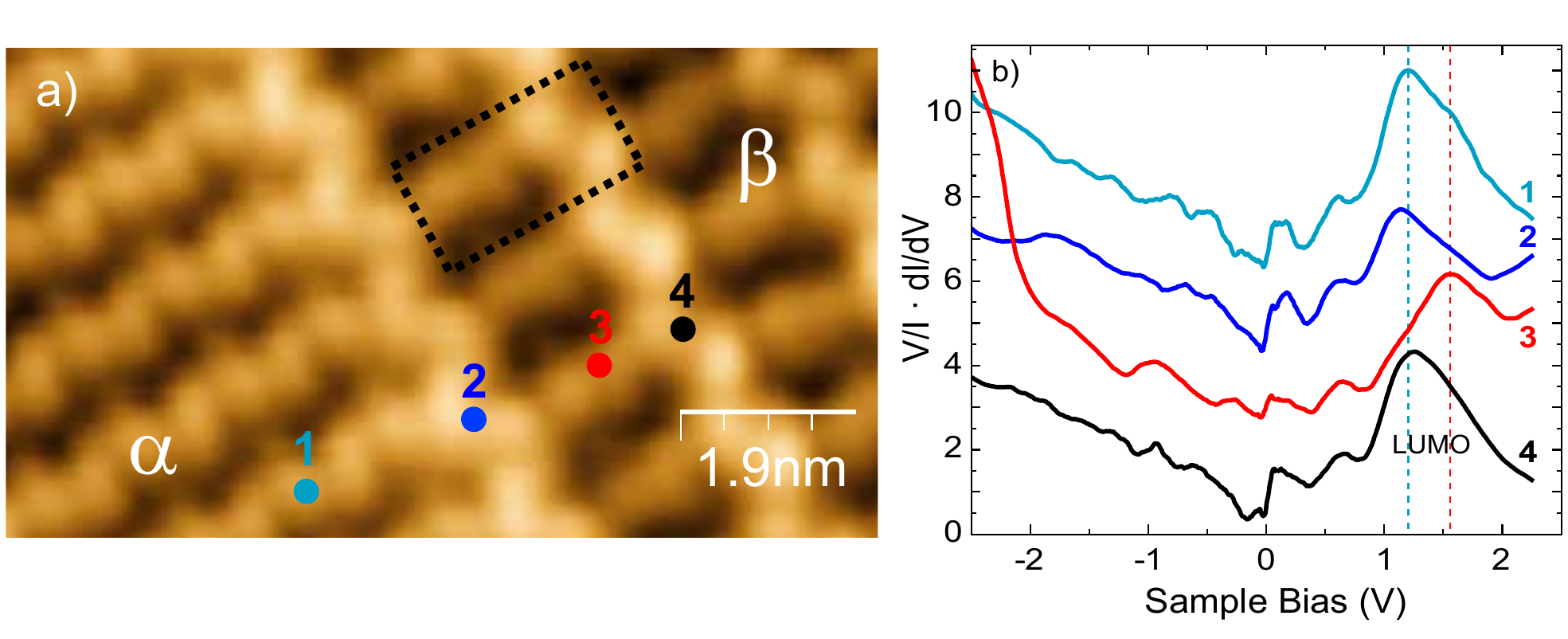}
  \end{center}
  \caption{(Color online) (a) STM image of transition region between molecular phases $\alpha$ and $\beta$.
  (Unit cell of phase $\beta$ marked as dotted box). (b) Normalized $\textrm{d}I/\textrm{d}V$ spectra on four different positions
  across the phases border, indicated in (a). The spectra 1, 2 and 4 are assigned to the SP species, while spectrum 3 shows
   significant deviations in both the LUMO position and the region below $-2$~eV. It is then assigned to the MC isomer.
  }
\end{figure}

At the center of the laser spot the photon fluence reaches its maximum value,
more than $3$~kJ/cm$^2$. There, all former patterns are transformed into a
single phase: the $\delta$ phase (the same as found after annealing a SP
monolayer to $330$~K (Fig. 1f). However, even after extended irradiation
periods of more than 8 hours, the $\delta$ phase was the only one observed; no
hint of the full MC phase (phase $\epsilon$ in Fig. 1) was ever found.
Following the tendency of an increase in the MC fraction with the irradiation
intensity, this structure is probably composed of a larger fraction of MC
isomers. The structure of the large unit cell appears rather inhomogeneous,
suggesting a varying ratio between the two isomers and/or involving various
conformers of the ''flexible'' merocyanine backbone. Unfortunately, our STM
measurements are not clear enough to reveal the precise isomer composition of
this phase.

The experiments thus show that illumination of a pure spiropyran monolayer on
Bi(110) results in pronounced changes in the molecular self-assembled
structures. The structural changes reflect the isomerization of large fractions
of the spiropyran molecules towards their merocyanine form. We can rule out a
thermal-activated cleavage of the C-O bond in spiropyran:  persistent
monitoring of the sample temperature and heat conduction calculations ensure
that the surface temperature does not rise significantly during the light
exposure. Thus, we attribute the isomerization to a photon-induced process.

The large-scale molecular reorganization implies that the two molecular species
are highly mobile on the surface during irradiation. Most probably the system
can be described as a two dimensional molecular ``liquid''; only in this way, a
gradual increase of the fraction of MC isomers with light would cause an
increase of MC-rich phases. The various mixed MC/SP structures are then
crystallized upon cooling. Their relative area changes gradually with the
position, reflecting the local composition of the isomeric mixture. At the
center of the laser spot, the mixed phase $\delta$ is always observed, even
after much larger illumination times (same light intensity but much larger
fluence). The stabilization of a MC:SP equilibrium structure independent of the
photon fluence is a fingerprint of the system being driven into a
photostationary state through SP~$\rightleftharpoons$~MC bidirectional
photo-isomerization processes. This contrasts with a situation where the
backswitching channel is blocked, what would inevitably lead to a saturation of
the pure reaction product MC \cite{note3}.

\begin{figure}[tb]
  \begin{center}
    \includegraphics[width=0.99\columnwidth]{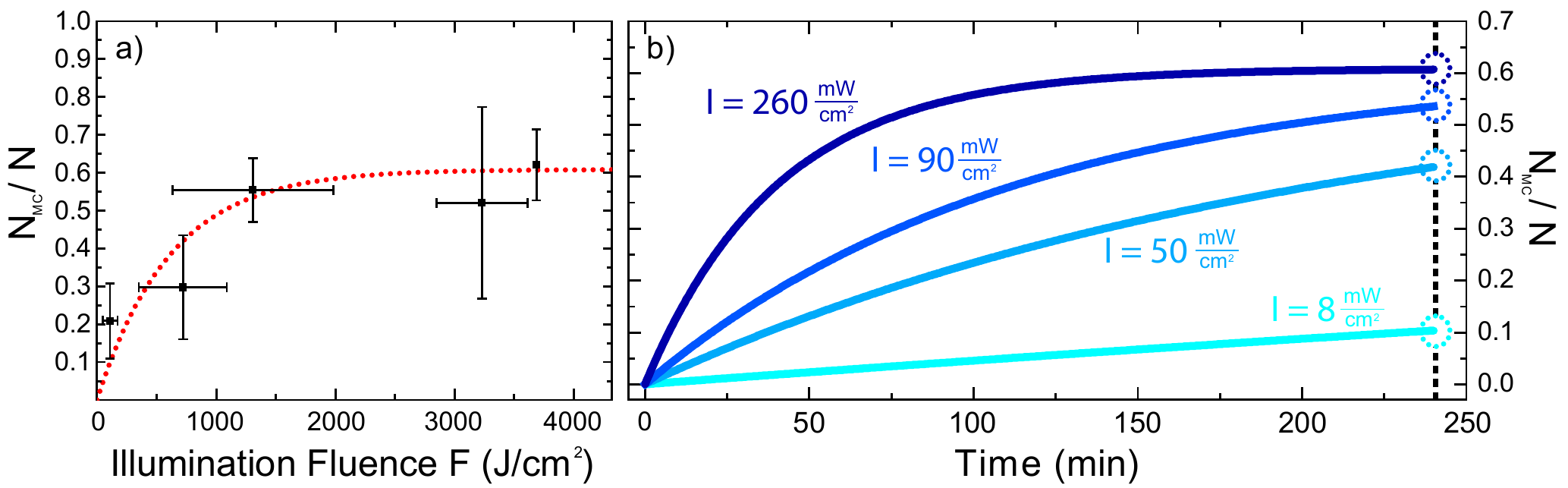}
  \end{center}
  \caption{(Color online) (a) Relative MC surface density on differently illuminated
  spots on the sample. Due to partially disordered regions in the inspected areas the error
   bars of the MC density is quite large. The fitted equilibration
  function (Eq. (2)) is represented by the red dotted curve. Each of the four curves
  shown in (b) represents
  one area with a certain illumination intensity as it was shown in Fig. 2 (a-d).
  The curves represent the respective amount of MC to SP molecules during the
  light irradiation procedure as it would result from function (Eq. (2)). The final
  values after 240 minutes of irradiation are marked with dotted circles.
  }
\end{figure}

In order to get a quantitative insight into the kinetics of the reaction, we
have estimated the fraction of MC isomers as a function of the position within
the laser spot. In each site, we used wide range STM images to determine the
relative area occupied by each phase and, using their MC composition obtained
above, the fraction of MC isomers in sample regions exposed to a certain
illumination. The resulting values are represented as a function of the fluence
as points in Fig. 4a. The saturation of the system into a photostationary mixed
phase is clear from these results.

We further analyze the observed behavior by fitting these data to a rate
equation model, typically used to describe equilibrium states in chemical
reactions. The number of MC isomers per unit of area, $n_\textrm{MC}$, follows
the rate equation:

\beqn \frac{\textrm{d}n_\textrm{MC}(t)}{\textrm{d}t}\;\;=\;\;P_\textrm{sm}
\cdot I \cdot n_\textrm{SP}(t)\;-\;P_\textrm{ms} \cdot I \cdot n_\textrm{MC}(t)
\eeqn%

where  $P_\textrm{sm}$ and $P_\textrm{ms}$ (in $($mJ$/$cm$^2)^{-1}$) are the
probabilities for photoexcitation of the SP~$\rightarrow$~MC and
MC~$\rightarrow$~SP reactions, respectively, $n_\textrm{SP}$ is the density of
SP isomers, and $I$ is the photon intensity. The quantities $n_\textrm{MC}$ and
$n_\textrm{SP}$ are treated as average densities for sufficiently large
microscopic areas. Density variations caused by diffusion through borders of
this area can be assumed to be negligible. Solving this differential equation
leads to the expression

\beqn
  \frac{n_\textrm{MC}(F)}{n_\textrm{MC}(F) + n_\textrm{SP}(F)}\;\;=\;\;
  \frac{1}{1+\frac{P_\textrm{ms}}
  {P_\textrm{sm}}}\;
  \left[ 1 - \exp\left\{-\:( P_\textrm{ms} +
  P_\textrm{sm} )\:\cdot\:F \right\} \right]
\eeqn

describing the fraction of MC molecules after irradiation with a given fluence
$F = I \cdot t$. A fit of the  data points of Fig. 4 (a) using the derived
function Eq. (2)  provides  an estimation of the photoexcitation rates of the
two isomerization processes: $P_\textrm{sm} = 9.9 \cdot 10^{-7} $~$($mJ $/$
cm$^2)^{-1}$ and $P_\textrm{ms} = 6.3 \cdot 10^{-7} $~$($mJ $/$ cm$^2)^{-1}$
\cite{note4}.

In our experiment, the different photon fluence in every sample position stems
from illuminating with different laser intensities during a fixed exposition
time. Using the results of the rate equations modelled above it is also
possible to monitor the time evolution of the fraction of the MC isomers in
sample areas exposed to  different laser intensities. The results, plotted in
Fig. 4 (b), show that the higher is the photon intensity in a region of the
laser spot, the earlier the photostationary state is reached. Furthermore,  the
whole sample would reach this state at large enough irradiation time scales.

From the obtained switching rates, we can derive the corresponding set of
photon cross sections  $\sigma_\textrm{sm} \approx 4 \cdot 10^{-22}$~cm$^2$ and
$\sigma_\textrm{ms} \approx 3 \cdot 10^{-22}$~cm$^2$ \cite{note5}. These values
are orders of magnitude lower than the typical photon cross sections of
SP~$\rightleftharpoons$~MC isomerization in solution, ranging between
$10^{-16}$~cm$^2$ and $10^{-15}$~cm$^2$ \cite{Goerner00}. On the other hand,
the efficiency on bismuth is at least two orders of magnitude larger than on a
gold surface, where the absence of observable switching events after similar
exposure as here imposes an upper limit to the cross section of
$10^{-24}$~cm$^2$ \cite{Tegeder11}. Both in gold and bismuth the employed
photon energy is smaller than the HOMO-LUMO energy difference, and larger than
either the HOMO or the LUMO alignment with respect to \Ef \cite{note6}. Since
an identical adsorption state is observed for SP layers on both surfaces,
similar cross-sections for photo-excitation are expected. The differences in
photo-isomerization activity are then ascribed not to the excitation process
itself, but to differences in the isomerization dynamics in the excited state.

Metal surfaces provide a continuum of electron-hole excitations that favors the
fast quenching of molecular excited states, thus drastically reducing the
excitation lifetime \cite{PerssonJPC78} and, consequently,  the switching
efficiency. Bismuth and gold differ in their density of electronic states in
the energy region around the Fermi level: bismuth, as a semimetal, has a lower
electron/hole density of states that could resonantly couple with the
photo-excited molecular resonances. The photo-excitations are then expected to
live longer, in agreement with the larger isomerization yields observed in our
experiments. In a similar way, photoactivity is also enhanced on insulating
surfaces or when the molecular species are functionalized with bulky endgroups
\cite{Hagen08a,Crommie07}; these decouple the molecule from the metal surface
and confine excitations in the molecule for larger time-scales. From our
results, the surface electronic structure thus appears as an additional crucial
parameter to steer the functionality of a molecular layer \cite{TrumpPRL04}.

A further intriguing aspect from the photon induced switching on bismuth is the
bidirectionality of the isomerization process with monochromatic light. In
fact, further experiments have shown that irradiation with photon energies from
the red-visible to the UV range lead to the $\delta$ phase as the final product
of the reaction. This apparent independence on photon-energy contrasts with the
behavior in  solution, where either of the two reactions can be activated  by
selecting light with photon energy matching  the corresponding absorption
transitions. We note that the photon energy of our experiments is too low to
induce a direct HOMO-LUMO transition. Instead, we consider that Bi(110) has a
narrow surface band around \Ef \cite{Hofmann} that is probably involved in the
excitation process. The continuum source of electronic states of this
substrate's band allows photon activated electron/hole transfer into molecular
states \cite{Hagen08b,Tegeder09}, supporting the bidirectional switching with a
single photon energy in a broad spectral range.

In summary, we have reported a photo-stationary ring-opening-closing state of
spiropyran on a bismuth surface. This state is based on the enhanced
photoactivity of these molecular switches  respect to other metal surfaces. We
attribute the origin of this enhancement to the the semimetallic character of
bismuth. The presence of the surface enables activation mechanisms using
photo-excited electrons from a surface band.

We thank C. Bronner, A. Kr\"uger, W. Kuch, F. Leyssner, and P. Tegeder for
fruitful discussions. Financial support by the DFG through Sfb 658 and SPP 1243
is gratefully acknowledged.


\begin{thebibliography}{99}
\bibitem{note1} Additional blank tests without illumination exclude that
    the   observed reactions were caused by external influences other than the
    laser illumination
\bibitem{note2} Although the size of the unit cell and appearance of the
    molecular features in the STM images suggests  a  MC/SP ratio  of 1:2 for
    phase $\beta$, we cannot completely exclude a 1:1 ratio, instead. The
    quantitative analysis described later in the text yields similar
    conclusions for both assumptions."
\bibitem{note3} We can discard that an eventual high thermodynamical
    stability of the $\delta$ mixed-phase could be the cause of its saturation
    upon irradiation because such stable phase should have been observed in more
    or less amount all over the sample. Besides, thermal full isomerization to MC films is observed
    at slightly higher temperatures.
\bibitem{note6}The HOMO-LUMO energy gap of SP in the $\alpha$ phase is
    $\approx$ 3.9 eV, following the results on Au(111) \cite{Tegeder11},
    where SP is physisorbed in an identical phase. For MC, the gap is $\approx$ 3.4 eV,
    based on the spectra of Fig. 3. Test measurements
    reproduced the reported isomerization with photon energies below 2.1 eV, supporting
    isomerisation through ionic transition states, rather than a HOMO-LUMO
    transition.
\end{thebibliography}
\end{document}